\documentclass[aps]{revtex4}
\usepackage[colorlinks=true, pdfstartview=FitV, linkcolor=blue, citecolor=red, urlcolor=magenta]{hyperref}
\usepackage{graphicx,color}  
\usepackage{amsfonts}
\begin{document}
\title{Analogue Aharonov-Bohm effect in a Lorentz-violating background}
\author{M. A. Anacleto, F. A. Brito and E. Passos}
\email{anacleto, fabrito, passos@df.ufcg.edu.br}
\affiliation{Departamento de F\'{\i}sica, Universidade Federal de Campina Grande, Caixa Postal 10071, 58109-970 Campina Grande, Para\'{\i}ba, Brazil}
\begin{abstract} 
In this paper we consider the acoustic black hole metrics obtained from a relativistic fluid under the influence of constant background that violates the Lorentz symmetry to study the analogue of the Aharonov-Bohm (AB) effect. We show that the scattering of planar waves by a draining bathtub vortex leads to a modified AB effect and due to the Lorentz symmetry breaking, the phase shift  persists even in the limit where the parameters associated
with the circulation and draining vanish. { In this limit, the Lorentz-violating background forms a conical defect, which is also responsible for the appearance of the analogue AB effect.}

\end{abstract}
\maketitle
\pretolerance10000

\section{Introduction}
The Aharonov-Bohm (AB) effect~\cite{Bohm} is one of the most extensively studied problems in planar
physics. This effect is essentially the scattering of charged particles
by a flux tube and has been experimentally confirmed~\cite{RGC}. In the quantum field theory the effect is simulated by a
nonrelativistic field theory describing spin zero particles
interacting through a Chern-Simons (CS) field~\cite{BL}. It was found to have analogues in gravitation \cite{FV}, fluid dynamics
\cite{CL}, optics \cite{NNK} and Bose-Einstein condensates \cite{LO}.

It was shown in~\cite{Berry} that scattering by a standard vortex leads
to an analogue of the AB effect, determined by a single dimensionless
circulation parameter $a$. Recently, it was shown in~\cite{Dolan} that the scattering of planar waves by a
'draining bathtub' vortex describes a modified AB effect
which depends on two dimensionless parameters associated with the circulation $a$ and draining $b$ rates.  The
effect is inherently asymmetric even in the low-frequency limit and leads to novel interference patterns.

The purpose of this paper is to investigate the effect of the Lorentz symmetry breaking on the scattering
by a 'draining bathtub' vortex which provides a simple
analogue for the AB effect in quantum mechanics. Thus, in this work we investigate 
how the AB effect due to a vortex flow is modified by the Lorentz symmetry breaking. As ours results show, we find that there appears small Lorentz violation correction to the scattering amplitude, which modify qualitative and quantitative
aspects of the AB effect. 

The noncommutative AB effect has been studied in the context of quantum mechanical
\cite{FGLR,Chai} and in the quantum field theory aproach~\cite{An}.
In~\cite{FGLR} the noncommutative AB effect has been
shown to be in contrast with the commutative situation. 
The cross section for the scattering of scalar particles by a
thin solenoid does not vanish even when the magnetic
field assumes certain discrete values. 

In the present calculations,  we apply the acoustic black hole metrics obtained from a relativistic fluid plus a
term (`a background field') that violates the Lorentz symmetry~\cite{ABP,ABP11} and we obtain a result  similar to the noncommutative AB effect \cite{FGLR}.
A relativistic version of acoustic black holes from the
noncommutative Abelian Higgs model has been also presented in~\cite{ABP12} (see also~\cite{Xian}).
It was found in~\cite{ABP11} that for suitable values of the Lorentz-violating parameter a wider
or narrower spectrum of particle wave function can be scattered with
increased amplitude by the acoustic black hole. This increases the
superresonance phenomenon previously studied in~\cite{SBP}. Thus, the presence of the Lorentz-violating parameter modifies
the quantity of removed energy of the acoustic black hole (see~\cite{MV,Volovik,Unruh} for some reviews).

In our study we shall focus on the differential cross section due to the scattering  of planar
waves by a draining bathtub vortex that leads to
a modified AB effect in a Lorentz-violating medium.  
We anticipate that we have obtained a cross section similar to that obtained in~\cite{FGLR} for noncommutative AB effect. The result implies that due to the Lorentz symmetry breaking pattern fringes
can still persist even in the limit where the parameters associated with the circulation $a$ and draining $b$ go to zero. { In this limit, the Lorentz-violating background forms a conical defect, which is also responsible for the appearance of the analogue AB effect.}

The paper is organized as follows. In Sec.~\ref{II} we apply the black hole metrics obtained in the extended Abelian Higgs model with the Lorentz-violating term~\cite{ABP,ABP11}. We then apply these metrics to a Klein-Gordon-like equation describing sound waves to study the scattering  of planar
waves by a draining bathtub vortex that leads to
a modified AB effect embedded into {\it two types} of a Lorentz-violating medium.   In Sec.~\ref{conclu} we make our final conclusions.

\section{The Lorentz-Violating Model}
\label{II}
In this section we shall apply the acoustic black hole metrics obtained in the extension of the Abelian Higgs model with a modified scalar sector via scalar Lorentz-violating term \cite{ABP,Bazeia:2005tb}.
The Lagrangian of the Lorentz-violating Abelian Higgs model is
\begin{eqnarray}
\label{acao}
{\cal L}&=&-\frac{1}{4}F_{\mu\nu}F^{\mu\nu} +|D_{\mu}\phi|^2+ m^2|\phi|^2-b|\phi|^4+ k^{\mu\nu}D_{\mu}\phi^{\ast}D_{\nu}\phi,
\end{eqnarray}
where $F_{\mu\nu}=\partial_{\mu}A_{\nu}-\partial_{\nu}A_{\mu}$, $D_{\mu}\phi=\partial_{\mu}\phi - ieA_{\mu}\phi$ and $k^{\mu\nu}$  is a constant tensor implementing the Lorentz symmetry breaking --- {a Lorentz-violating background}. The tensor coefficient is a $4\times4$ matrix, given by
\begin{equation}
k_{\mu\nu}=\left[\begin{array}{clcl}
\beta &\alpha &\alpha & \alpha\\
\alpha &\beta &\alpha &\alpha \\
\alpha &\alpha &\beta &\alpha\\
\alpha &\alpha &\alpha &\beta
\end{array}\right], \quad(\mu,\nu=0,1,2,3),
\end{equation}
with $\alpha$ and $\beta$ being real parameters. In a previous study \cite{ABP} following this theory we have found three- and two-dimensional acoustic metric describing acoustic black holes. In the following we shall focus on the planar acoustic black hole metrics to address the issue of analogue Aharonov-Bohm effect. { We shall work by restricting the full set of parameters to the cases $\beta\neq0$, $\alpha=0$ and $\beta=0$, $\alpha\neq0$. We decide for doing this, because these two choices have shown to be rich enough to develop our study, whereas considering the full set of parameters turns the analysis very hard and not satisfactorily illuminating.

Let us briefly review the steps to find acoustic metrics from quantum field theory. Firstly, we decompose the scalar field as  $\phi = \sqrt{\rho(x, t)} \exp {(iS(x, t))}$ into the original Lagrangian to find
\begin{eqnarray}
&&{\cal L} = -\frac14 F_{\mu\nu}F^{\mu\nu} + \rho\partial_{\mu}S\partial^{\mu}S - 2e\rho A_\mu \partial^{\mu}S+ e^2\rho A_\mu A^\mu + m^2\rho - b\rho^2\nonumber\\
&+& k^{\mu\nu} \rho(\partial_{\mu}S\partial_{\nu}S-2eA_\mu\partial_{\nu}S+eA_\mu A_\nu)
+\frac{\rho}{\sqrt{\rho}}(\partial_{\mu}\partial^{\mu}+k^{\mu\nu}\partial_{\mu}\partial_{\nu})\sqrt{\rho}.
\end{eqnarray}
Secondly, linearizing the equations of motion around the background $(\rho_0,S_0)$, with $\rho=\rho_0+\rho_1$ and  $S=S_0+\psi$ we find the equation of motion for a linear acoustic disturbance $\psi$ given by a Klein-Gordon equation in a curved space
\begin{eqnarray}
\frac{1}{\sqrt{-g}}\partial_{\mu}(\sqrt{-g}g^{\mu\nu}\partial_{\nu})\psi=0,
\end{eqnarray}
where $g_{\mu\nu}$ just represents the acoustic metrics given explicitly in the examples below.
 
}

\pretolerance10000

\subsection{The case $\beta\neq0$ and $\alpha=0$}
The acoustic line element in polar coordinates on the plane with Lorentz symmetry breaking, up to an irrelevant position-independent factor, in the `non-relativistic' limit ($v^2\ll c^2$) is given by \cite{ABP}
\begin{eqnarray}
ds^2=-\frac{(c^2-{\tilde\beta}_{-}v^2)}{{\tilde\beta}_{+}}dt^2-2(v_{r}dr+v_{\phi}rd\phi)dt+\frac{{\tilde\beta}_{+}}{{\tilde\beta}_{-}}(dr^2+r^2d\phi^2),
\end{eqnarray}
where $\tilde{\beta}_{\pm}\equiv 1\pm\beta$, $c$ is the sound velocity in the fluid and $v$ is the fluid velocity.
We consider the flow with the velocity potential $\psi(r,\phi) = A\ln{r} + B\phi$  whose velocity profile in polar coordinates on the plane is  given by
\begin{eqnarray}
\vec{v}=\frac{A}{r}\hat{r}+\frac{B}{r}\hat{\phi}.
\end{eqnarray}
{Let us now consider the} transformations of the time and the azimuthal angle coordinates as follows
\begin{eqnarray}
d\tau&=&dt+\frac{{\tilde\beta}_{+}Ardr}{(c^2r^2-{\tilde\beta}_{-}A^2)},
\\
d\varphi&=&d\phi+\frac{B{\tilde\beta}_{-}Adr}{r(c^2r^2-{\tilde\beta}_{-}A^2)}.
\end{eqnarray}
In these new coordinates the metric becomes
\begin{eqnarray}
\label{ELB}
ds^2\!=\!\frac{{\tilde\beta}_{+}}{{\tilde\beta}_{-}}
\left[-\frac{{\tilde\beta}_{-}}{{\tilde\beta}_{+}^2}\left(1-\frac{{\tilde\beta}_{-}(A^2+B^2)}{c^2r^2}\right)d\tau^2
+\left(1-\frac{{\tilde\beta}_{-}A^2}{c^2r^2}\right)^{-1}dr^2
-\frac{2{\tilde\beta}_{-}B}{{\tilde\beta}_{+}c}d\varphi d\tau+r^2d\varphi^2\right].
\end{eqnarray}
The radius of the ergosphere is given by $g_{00}(r_{e}) = 0$, whereas the horizon is given by the coordinate singularity $g_{rr}(r_{h}) = 0$, that is
\begin{eqnarray}
r_{e}=\sqrt{r_{h}^2+\frac{\tilde\beta_{-}B^2}{c^2}}, \quad r_{h}=\frac{{\tilde\beta}_{-}^{1/2}|A|}{c}.
\end{eqnarray}
We can observe from the equation (\ref{ELB}) that for $A > 0$ we are dealing
with a past event horizon, i.e., acoustic white hole and
for $A < 0$ we are dealing with a future acoustic horizon, i.e.,
acoustic black hole.
{ The metric can be now written in the form}
\begin{eqnarray}
g_{\mu\nu}=\frac{{\tilde\beta}_{+}}{{\tilde\beta}_{-}}\left[\begin{array}{clcl}
-\frac{{\tilde\beta}_{-}}{{\tilde\beta}_{+}^2}\left[1-\frac{r_{e}^2}{r^2}\right] &\quad\quad 0& -\frac{{\tilde\beta}_{-}}{{\tilde\beta}_{+}}\frac{B}{cr}\\
0 & \left(1-\frac{r_{h}^2}{r^2} \right)^{-1}& 0\\
-\frac{{\tilde\beta}_{-}}{{\tilde\beta}_{+}}\frac{B}{cr} &\quad\quad 0 & 1
\end{array}\right],
\end{eqnarray}
with inverse $g^{\mu\nu}$
\begin{eqnarray}
\label{metrinv}
g^{\mu\nu}=\frac{{\tilde\beta}_{+}}{{\tilde\beta}_{-}}\left[\begin{array}{clcl}
-\frac{\tilde{\beta}_{+}^2}{\tilde{\beta}_{-}}f(r)^{-1} &\quad\quad 0& -\frac{\tilde{\beta}_{+}B}{cr}f(r)^{-1}\\
0 & \left(1-\frac{r_{h}^2}{r^2} \right)& 0\\
-\frac{\tilde{\beta}_{+}B}{cr}f(r)^{-1} &\quad\quad 0 & \left[1-\frac{r_{e}^2}{r^2}\right]f(r)^{-1}
\end{array}\right],
\end{eqnarray}
where $f(r)=1-\frac{r_{h}^2}{r^2}$. 

{We shall now consider} the Klein-Gordon equation for a linear acoustic disturbance $\psi(t,r,\phi)$ in the background metric (\ref{metrinv}), i.e.,
\begin{eqnarray}
\frac{1}{\sqrt{-g}}\partial_{\mu}(\sqrt{-g}g^{\mu\nu}\partial_{\nu})\psi=0.
\end{eqnarray}
We can make a separation of variables into the equation above as follows
\begin{eqnarray}
\psi(t,r,\phi)=R(r)e^{i(\omega t-m\phi)}.
\end{eqnarray}
The radial function $R(r)$ satisfies the linear second order differential equation 
\begin{eqnarray}
\label{EQKG}
&&\left[\frac{\tilde{\beta}_{+}^2}{\tilde{\beta}_{-}}\omega^2-\frac{2\tilde{\beta}_{+}Bm\omega}{cr^2}-\frac{m^2}{r^2}\left(1-\frac{r_{e}^2}{r^2}\right)\right]R(r)
+\frac{f(r)}{r}\frac{d}{dr}\left[rf(r)\frac{d}{dr}\right]R(r)=0.
\end{eqnarray} 
We now introduce the tortoise coordinate $r^{\ast}$ by using the following equation
\begin{eqnarray}
\label{tc}
\frac{d}{dr^{\ast}}=f(r)\frac{d}{dr}, \quad f(r)=1-\frac{r_{h}^2}{r^2}=1-\frac{{\tilde\beta}_{-}A^2}{c^2r^2},
\end{eqnarray}
which gives the solution
\begin{eqnarray}
r^{\ast}=r+\frac{\sqrt{{\tilde\beta}_{-}}|A|}{2c}\log{\left(\frac{r-\frac{\sqrt{{\tilde\beta}_{-}}|A|}{c}}{r+\frac{\sqrt{{\tilde\beta}_{-}}|A|}{c}}\right)}.
\end{eqnarray}
Observe that in this new coordinate the horizon $r_{h}=\frac{{\tilde\beta}_{-}^{1/2}|A|}{c}$
maps to $r^{\ast}\rightarrow-\infty$ 
while $r\rightarrow\infty$ corresponds to $r^{\ast}\rightarrow+\infty$.
Now, we consider a new radial function, $G(r^{\ast})=r^{1/2}R(r)$ and the modified radial equation obtained from (\ref{EQKG}) is
\begin{eqnarray}
\label{EG}
&&\frac{d^2G(r^{\ast})}{dr^{\ast2}}
+\left[\left(\frac{\tilde{\beta}_{+}}{\tilde{\beta}_{-}^{1/2}}\omega-\frac{\tilde{\beta}_{-}^{1/2}Bm}{cr^2}\right)^2-V(r)\right]G(r^{\ast})=0, 
\end{eqnarray}
where $V(r)$ is the potential given by
\begin{equation}
V(r)=\frac{f(r)}{4r^2}\left(4m^2-1+\frac{5\tilde{\beta}_{-}A^2}{c^2r^2}\right),
\end{equation}
{a form that resembles that given in Ref.~\cite{Dolan}.}
In the asymptotic region ($r^{\ast}\rightarrow\infty$), 
we find for the equation (\ref{EG}) the simple solution
\begin{eqnarray}
\label{sl}
G(r^{\ast})={\cal C}_{m}e^{i\tilde{\omega}r^{\ast}}+{\cal R}_{m}e^{-i\tilde{\omega}r^{\ast}},
\quad \tilde{\omega}=\frac{\tilde{\beta}_{+}}{\tilde{\beta}_{-}^{1/2}}\omega.
\end{eqnarray}
Notice that the first term in equation (\ref{sl}) corresponds to ingoing wave and the second term to the reflected wave, so that ${\cal R}$ is the reflection coefficient, given by
\begin{equation}
{\cal R}_{m}=\frac{i^{1/2}(-1)^m}{\sqrt{2\pi\omega}}.
\end{equation}
Now, near the horizon region ($r^{\ast}\rightarrow-\infty$), we have
\begin{eqnarray}
G(r^{\ast})={\cal T}_{m}e^{i\left(\tilde{\omega}-m\tilde{\Omega}_{H}\right)r^{\ast}},
\end{eqnarray}
where, $\tilde{\Omega}_{H}=\Omega_{H}/\sqrt{\tilde{\beta}_{-}}$ and $\Omega_{H}=Bc/A^2$ is the angular velocity of the acoustic black hole and 
${\cal T}$ is the transmission coefficient.

\subsubsection{Analogue Aharonov-Bohm Effect}

Let us now consider the analogue Aharonov-Bohm effect by considering the scattering of a monochromatic planar
wave of frequency $\omega$ given in the form
\begin{eqnarray}
\psi(t,r,\phi)=e^{-i\omega t}\sum_{m=-\infty}^{\infty}R_{m}(r) e^{im\phi}/\sqrt{r},
\end{eqnarray}
such that far from the vortex, the function $\psi$ can be written in terms of the sum of a plane wave and a scattered wave, i.e., 
\begin{eqnarray}
\psi(t,r,\phi)\sim e^{-i\omega t}(e^{i\omega x}+f_{\omega}(\phi)e^{i\omega r}/\sqrt{r}),
\end{eqnarray}
where $e^{i\omega x}=\sum_{m=-\infty}^{\infty}i^mJ_{m}(\omega r) e^{im\phi}$ and $J_{m}(\omega r)$ 
is a Bessel function of the first kind. { The scattering amplitude $f_{\omega}(\phi)$ is given in the
partial-wave form}
\begin{eqnarray}
f_{\omega}(\phi)= \sqrt{\frac{1}{2i\pi\omega}}\sum_{m=-\infty}^{\infty}(e^{2i\delta_{m}}-1) e^{im\phi},
\end{eqnarray}
{with the phase shift given by}
\begin{equation}
e^{2i\delta_{m}}=i(-1)^m\frac{{\cal C}}{{\cal R}}.
\end{equation}

{In order to compute the phase shift, at some level of approximation, let us first rewrite the equation (\ref{EG}) in terms of a new function $X(r)=f(r)^{1/2}G(r^{\ast})$, that is}
\begin{eqnarray}
&&\frac{d^2X(r)}{dr^{2}}+ \left(\frac{3r_{h}^2}{f(r)r^4}+\frac{r_{h}^4}{f(r)^2r^6}\right)X(r)
+\left[\left(\frac{\tilde{\beta}_{+}}{\tilde{\beta}_{-}^{1/2}}\omega-\frac{\tilde{\beta}_{-}^{1/2}Bm}{r^2}\right)^2-V(r)\right]
\frac{X(r)}{f^2(r)}=0.
\end{eqnarray}
{Now performing a power series in $1/r$,  this equation can be written as}
\begin{eqnarray}
\frac{d^2X(r)}{dr^{2}}
+\left[\tilde{\omega}^2-\frac{(4\tilde{m}^2-1)}{4r^2}-U(r)\right]X(r)=0, 
\end{eqnarray}
where $\tilde{m}^2=m^2+2am-2b^2$, $a=\tilde{\omega}\tilde{\beta}_{-}^{1/2}B$, $b=\tilde{\omega}\tilde{\beta}_{-}^{1/2}A$ and,
\begin{eqnarray}
U(r)=\frac{(a^2-b^2)m^2-4b^2am+2b^2+3b^4}{\tilde{\omega}^2r^4}+\frac{b^2(2a^2-b^2)m^2-6b^4am+3b^4+4b^6}{\tilde{\omega}^4r^6}
+O(\tilde{\omega}^{-6}r^{-8}),
\end{eqnarray}
being $a$ and $b$ parameters that describe the coupling to circulation and draining, respectively. Now applying the approximation formula
\begin{eqnarray}
\delta_{m}\approx \frac{\pi}{2}(m-\tilde{m})+\frac{\pi}{2}\int^{\infty}_{0}r[J_{\tilde{m}}(\widetilde{\omega}r)]^2U(r)dr,
\end{eqnarray}
and using $|m|\gg\sqrt{a^2+b^2}$, we obtain~\cite{Dolan}
\begin{eqnarray}
\label{fase}
\delta_{m}\cong- \frac{a\pi}{2}\frac{m}{|m|}+\frac{3\pi(a^2+b^2)}{8|m|}-\frac{5a\pi(a^2+b^2)}{8m^2}\frac{m}{|m|}.
\end{eqnarray}
Note that the above result for the mode $m=0$ is not valid but in the limits $m\rightarrow \pm \infty$ the first term in (\ref{fase}) implies that the phase shifts tend
to non-zero constants, which naturally leads 
to an AB effect~\cite{Fetter}. { Furthermore, the isotropic mode $m = 0$, that we can obtain from equation (\ref{EG})}
\begin{eqnarray}
G_{m=0}(r^{\ast})=r^{1/2}e^{b\pi/2}J_{ib}(\omega rf^{1/2}),
\end{eqnarray}
and the phase shift is imaginary
\begin{eqnarray}
\label{fase2}
\delta_{m=0}=\frac{1}{2}i\pi b.
\end{eqnarray}
Because $a, b$ are proportional to $\tilde{\omega}$ and $\tilde{\beta}_\pm$, we can conclude from equations (\ref{fase}) and (\ref{fase2}) that the AB effect is dominant in the scattering of low-frequencies waves, ${\omega}\sqrt{A^2+B^2}\ll1$ or high-frequencies waves,  ${\omega}\tilde{\beta}_+\sqrt{A^2+B^2}\ll1$ in the absence or presence of a `strong' Lorentz-violating background with $\tilde{\beta}_+\simeq0$, respectively.

Thus, by using the eqs. (\ref{fase}) and (\ref{fase2}), to lowest order
in $a$, $b$, we can compute the differential scattering cross section (with units of length) that is given by
\begin{eqnarray}
\frac{d\sigma_{ab}}{d\phi}=|f_{\omega}(\phi)|^2\cong \frac{\pi}{2\tilde{\omega}}\frac{[a\cos(\phi/2)-b\sin(\phi/2)]^2}{\sin^{2}(\phi/2)}.
\end{eqnarray}
For $b=0$ (the non-draining limit), we have the vortex result of Fetter~\cite{Fetter} 
\begin{eqnarray}
\label{eqvortex}
\frac{d\sigma_{vortex}}{d\phi}= \frac{\pi a^2}{2\tilde{\omega}}\cot^2(\phi/2)=\frac{(1-\beta)^{1/2}\pi^2 a^2}{2\pi{(1+\beta)}\omega}\cot^2(\phi/2),
\end{eqnarray}
which also resembles the exact Aharonov-Bohm effect \cite{Bohm} 
\begin{eqnarray}
\frac{d\sigma_{AB}}{d\phi}=\frac{1}{2\pi\tilde{\omega}}\frac{\sin^2(\pi a)}{\sin^{2}(\phi/2)}
=\frac{(1-\beta)^{1/2}}{2\pi{(1+\beta)}\omega}\frac{\sin^2(\pi a)}{\sin^{2}(\phi/2)},
\end{eqnarray}
for small angle or small coupling limits. Particularly,  for small Lorentz-symmetry breaking parameter $\beta$ we find
\begin{eqnarray}
\frac{d\sigma_{AB}}{d\phi}=\frac{(1-3\beta/2)}{2\pi\omega}\frac{\sin^2(\pi a)}{\sin^{2}(\phi/2)}.
\end{eqnarray}
In addition, for small $\beta$ and small angle $\phi$, the equation (\ref{eqvortex}) becomes
\begin{eqnarray}
\frac{d\sigma_{vortex}}{d\phi}=\frac{(1-3\beta/2)\pi^2 a^2}{2\pi\omega}\left[\frac{4}{\phi^2}-\frac{2}{3}+\frac{\phi^2}{60}+O(\phi^3)\right].
\end{eqnarray}
These results show that our setup present corrections which modify the qualitative and quantitative aspects of the AB effect.

\subsection{The case $\beta=0$ and $\alpha\neq0$}
In the present subsection we repeat the previous analysis for $\beta=0$ and $\alpha\neq0$. As in the earlier case we take the acoustic line element with Lorentz symmetry breaking obtained in  \cite{ABP}  in the non-relativistic limit, up to first order in $\alpha$, given by
\begin{eqnarray}
\label{am}
ds^2=-\tilde{\alpha}\left(1-\frac{r_{e}^2}{r^2}\right)d\tau^2
+\tilde{\alpha}^{-1}\left(1-\frac{r_{h}^2}{r^2}\right)^{-1}dr^2
-\frac{2B}{cr}rd\varphi d\tau+\left[1+\frac{2\alpha(\tilde{\alpha}^{1/2}cr_{h}+B)}{r}\right]r^2d\varphi^2,
\end{eqnarray}
where $\tilde{\alpha}=1+\alpha$. 
The radius of the ergosphere $(r_{e})$ and the horizon $(r_{h})$ { now reads}
\begin{eqnarray}
r_{e}=\sqrt{r_{h}^2+\frac{B^2}{\tilde{\alpha}c^2}}, \quad r_{h}=\frac{|A|}{\tilde{\alpha}^{1/2}c}.
\end{eqnarray}
Now, the radial function $R(r)$, as in the previous case, satisfies the linear second order differential equation
\begin{eqnarray}
\label{EKG}
\left[\omega^2-\frac{2Bm\omega\eta(r)}{cr^2}
-\frac{\tilde{\alpha}m^2\eta(r)}{r^2}\left(1-\frac{r_{e}^2}{r^2}\right)\right]\frac{f(r)R(r)}{{\cal D}(r)}
%\nonumber\\
+\frac{\tilde{\alpha}f(r)}{r}\frac{d}{dr}\left[rf(r)\frac{d}{dr}\right]R(r)=0,
\end{eqnarray}
where $\eta(r)=\left[1+\frac{2\alpha(\tilde{\alpha}^{1/2}cr_{h}+B)}{r}\right]^{-1}$ 
and ${\cal D}(r)=\frac{B^2\eta(r)}{c^2r^2}+\tilde{\alpha}
[1-\frac{B^2}{\tilde{\alpha}c^2r^2}]$. 
Again, we introduce the tortoise coordinate $r^{\ast}$ through the equation (\ref{tc})
%$\frac{d}{dr^{\ast}}=f(r)\frac{d}{dr}$, 
and after introducing a new radial function, $G(r^{\ast})=r^{1/2}R(r)$, the modified radial equation (\ref{EKG}), becomes
\begin{eqnarray}
\label{EKGmod}
\frac{d^2G(r^{\ast})}{dr^{\ast 2}}+
\left[\omega^2-\frac{2Bm\omega\eta(r)}{cr^2}
-\frac{m^2\eta(r)B^2}{c^2r^4}\right]\frac{f(r)G(r^{\ast})}{\tilde{\alpha}{\cal D}(r)}-V(r)G(r^{\ast})=0,
%\nonumber\\
\end{eqnarray}
where
\begin{equation}
V(r)=\frac{f(r)}{4r^2}\left(\frac{4m^2\eta(r)f(r)}{{\cal D}(r)}-1+\frac{5A^2}{\tilde{\alpha}c^2r^2}\right).
\end{equation}

\subsubsection{Analogue Aharonov-Bohm Effect}

{ As in the previous case, in order to compute the phase shift, at some level of approximation, let us first rewrite the equation (\ref{EKGmod}) in terms of a new function $X(r)=f(r)^{1/2}G(r^{\ast})$, that is
\begin{eqnarray}\label{ekg-m}
\frac{d^2X(r)}{dr^{2}}+ \left[\frac{3r_{h}^2}{f(r)r^4}+\frac{r_{h}^4}{f(r)^2r^6}+
\left(\omega^2-\frac{2Bm\omega\eta(r)}{cr^2}
-\frac{m^2\eta(r)B^2}{c^2r^4}\right)\frac{1}{\tilde{\alpha}{\cal D}(r)f(r)}-\frac{V(r)}{f^2(r)}\right]X(r)=0.
%\nonumber\\
\end{eqnarray}
Now, by power expanding in powers of $1/r$, the equation (\ref{ekg-m}) becomes
\begin{eqnarray}
\frac{d^2X(r)}{dr^{2}}
+\left[\tilde{\omega}^2-\frac{(4\tilde{m}^2-1)}{4r^2}+\frac{4\alpha\tilde{\alpha}^{1/2}m(a+b)a^3}{\tilde{\omega}r^2}-U(r)-{\cal{V}}(r)\right]X(r)=0, 
\end{eqnarray}
where we define the parameters 
%\[
$\tilde{m}^2=m^2/\tilde{\alpha}+2am-2b^2, \quad a=\tilde{\omega}B/\tilde{\alpha}^{1/2}, \quad b=\tilde{\omega}A/\tilde{\alpha}^{1/2}, 
\quad \tilde{\omega}=\omega/\tilde{\alpha},$ and the functions as }
%\]
\begin{eqnarray}
U(r)=\frac{(a^2-b^2)m^2-4b^2am+2b^2+3b^4}{\tilde{\omega}^2r^4}+\frac{b^2(2a^2-b^2)m^2-6b^4am+3b^4+4b^6}{\tilde{\omega}^4r^6}
+O(\tilde{\omega}^{-6}r^{-8}),
\end{eqnarray}
and 
\begin{eqnarray}
{\cal{V}}(r)&=&2\alpha\tilde{\alpha}^{1/2}(a+b)\left[\frac{a^2-m^2}{\tilde{\alpha}\tilde{\omega}r^3}
-\frac{2ma^3b^2}{\tilde{\omega}^3r^4}
+\frac{a^2[b^2+m^2(1+a^2\tilde{\alpha})-2ma]}{\tilde{\alpha}\tilde{\omega}^3r^5}\right.
\nonumber\\
&+&\left.\frac{ma^4-(2ma^3-m^2a^4\tilde{\alpha})b^2}{\tilde{\alpha}\tilde{\omega}^5r^7}\right]+\cdots
\end{eqnarray}
{As in the previous case, by applying again the approximation formula
\begin{eqnarray}
\delta_{m}\approx \frac{\pi}{2}(m-\tilde{m})+\frac{\pi}{2}\int^{\infty}_{0}r[J_{\tilde{m}}(\widetilde{\omega}r)]^2[U(r)+{\cal{V}}(r)]dr,
\end{eqnarray}
we readily obtain}
\begin{eqnarray}
\delta_{m}&\cong&- \sqrt{{\tilde{\alpha}}}\left(\frac{a\pi}{2}+\alpha(a+b)\right)\frac{m}{|m|}
+\frac{\pi(3a^2{\tilde{\alpha}}^{3/2}-b^2({\tilde{\alpha}}-4){\tilde{\alpha}}^{1/2})+16\alpha{\tilde{\alpha}}^{3/2}(a^2+ab)}{8|m|}+
\nonumber\\
&&-\frac{a\pi(5a^2{\tilde{\alpha}}^{5/2}-b^2(3{\tilde{\alpha}}-8){\tilde{\alpha}}^{3/2})}{8m^2}\frac{m}{|m|}
+\frac{8\alpha^2{\tilde{\alpha}}^2(a^5+a^4 b)-6\pi a^4b^2\alpha{\tilde{\alpha}}^{2}}{12m^2}\frac{m}{|m|}
\nonumber\\
&&+\frac{a^3\alpha{\tilde{\alpha}^{1/2}}
[(4(2-12{\tilde{\alpha}}){\tilde{\alpha}}+3)-6\pi b^3{\tilde{\alpha}}]}{12m^2}\frac{m}{|m|}
+\frac{a^2\alpha{\tilde{\alpha}^{1/2}}(4(2-12{\tilde{\alpha}})
{\tilde{\alpha}}+3)}{12m^2}\frac{m}{|m|}
\nonumber\\
&&-\frac{\alpha{\tilde{\alpha}}^{3/2}[a(24b^2+3)+3b(8b^2+1)]}{12m^2}\frac{m}{|m|}
+\frac{\pi}{2}(1-{\tilde{\alpha}}^{-1/2})m\frac{m}{|m|}.
\end{eqnarray}
Thus, to lowest order
in $a$, the differential scattering cross section with $b=0$ and for small $\alpha$, is
\begin{eqnarray}
\frac{d\sigma_{vortex}}{d\phi}&=&|f_{\omega}(\phi)|^2\cong 
\frac{1}{2\pi {\tilde{\omega}}}\left|\sum_{m=-\infty}^{\infty}
(e^{-2i\tilde{a}+i\pi\lambda m}-1)e^{im\phi}\right|^2,
\end{eqnarray}
where, $\tilde{a}=a\sqrt{{\tilde{\alpha}}}\left(\frac{\pi}{2}+\alpha\right)$ and  $\lambda=1-{\tilde{\alpha}}^{-1/2}$. For small angles $\phi\neq 0$, we have
\begin{eqnarray}
\frac{d\sigma_{vortex}}{d\phi}
&=&\frac{(1+\alpha)}{2\pi \omega}
\left[\left(\frac{16-4\pi^2\lambda^2/3}{\phi^2}+\frac{4-3\pi^2}{3}-\frac{4}{45}\pi^2\lambda^2+\phi^2\left(\frac{1}{15}-\frac{\pi^2\lambda^2}{3780}\right)\right)\tilde{a}^2\right.
\nonumber\\
&+&\left.\left(\frac{2\pi^2}{3\phi^2}+\frac{11\pi^2}{180}+\frac{31\pi^2\phi^2}{7560}\right)\lambda^2\right]+O(\phi^3,\tilde{a}^3, \lambda^3)
+less\,\, singular\,\, terms.
\end{eqnarray}
{ The dominant term in the differential scattering cross section is $1/\phi^2$}
\begin{eqnarray}
\frac{d\sigma_{vortex}}{d\phi}&=&\frac{(1+\alpha)}{2\pi \omega}
\left[\left(\frac{16-4\pi^2\lambda^2/3}{\phi^2}\right)\tilde{a}^2
+\frac{2\pi^2}{3\phi^2}\lambda^2\right]+less\,\, singular\,\, terms
\nonumber\\
&=&\frac{(1+\alpha)}{2\pi \omega}
\left[\left(\frac{16-2\pi^2\alpha^2/3}{\phi^2}\right)\frac{\pi^2{a}^2}{4}\left(1+\frac{4\alpha}{\pi}\right)
+\frac{\pi^2}{3\phi^2}\alpha^2\right]+less\,\, singular\,\, terms.
\end{eqnarray}

Now, if the $a=0$, the differential cross section at small angles is
dominated by
\begin{eqnarray}
\frac{d\sigma_{vortex}}{d\phi}
&=&\alpha^2\frac{\pi}{6\omega\phi^2}+less\,\, singular\,\, terms.
\end{eqnarray}
Note that, contrarily to the usual Aharonov-Bohm effect, in the case with Lorentz symmetry breaking the differential
scattering cross section is different from zero when $a=0$. 
Our results are qualitatively in agreement with that obtained in~\cite{FGLR}, for the AB effect in the context of non-commutative quantum mechanics. This correction vanishes in the limit $\alpha\rightarrow 0$  so that no singularities are generated.
This correction ($\alpha^2$) due to effect of Lorentz symmetry breaking may be relevant at high energies.
 Our result shows that pattern fringes can appear even when $a=0$, unlike the usual case.
 
{ One can understand this effect as follows. In the limit of circulation $a=\tilde{\omega}B/\tilde{\alpha}^{1/2}$ and draining  $b=\tilde{\omega}A/\tilde{\alpha}^{1/2}$ vanishes then for nonzero $\tilde{\omega}=\omega/\tilde{\alpha}$ and finite $\tilde{\alpha}=1+\alpha$, we automatically have $A=B=r_h=r_e=0$ such that the metric (\ref{am}) simply becomes the metric of a conical defect 
 \begin{eqnarray}
\label{am-2}
\tilde{\alpha}ds^2=-\tilde{\alpha}^2d\tau^2
+dr^2
+r^2\tilde{\alpha}d\varphi^2,
\end{eqnarray}
with angle deficit $\delta=2\pi(1-\sqrt{\tilde{\alpha}})$.
 Thus, even though there is no vortex in the above limit, the Lorentz-violating background forms a conical defect, which is responsible for the appearance of the analogue AB effect in a even more faithfully way. 
 }
\section{Conclusions}
\label{conclu}

In this paper we have considered the implications of a Lorentz-violating background on the analogue Aharonov-Bohm effect. We have used the acoustic metric obtained in an extended Abelian Higgs model with a Lorentz-violating term \cite{ABP}. The results have shown that unlike the usual Aharonov-Bohm effect, its analogue has a major distinct feature, that is, the fact of the differential
scattering cross section being different from zero even if the parameters that controls the circulation and draining become zero. Another interesting phenomena that appeared in this setup is that the scattering of low or high-frequencies waves is directly affected by the parameter that controls the Lorentz symmetry breaking.

\acknowledgments

We would like to thank CNPq, CAPES, PNPD/PROCAD -
CAPES for partial financial support.

\end{document}